\documentclass[structabstract]{aa}

\usepackage{amssymb}
\usepackage{amsmath}
\usepackage{graphicx}
\usepackage{url}

\usepackage[utf8]{inputenc}
\usepackage{natbib}

\begin{document}
   \title{Improved angular momentum evolution model for solar-like stars}


   \author{F. Gallet          
          \and
          J. Bouvier
          }

   \institute{UJF-Grenoble 1/CNRS-INSU, Institut de Plan\'etologie et d'Astrophysique de Grenoble (IPAG) UMR 5274,
   F-38041 Grenoble, France\\
   \email{florian.gallet@obs.ujf-grenoble.fr}}

   \date{Received 15 February 2013; accepted 6 June 2013} 

 \abstract{Understanding the origin and evolution of stellar angular momentum is one of the major challenges of stellar physics.}
 {We present new models for the rotational evolution of solar-like stars between 1~Myr and 10~Gyr with the aim of reproducing the distributions of rotational periods observed for star forming regions and young open clusters within this age range.}{The models include a new wind braking law based on recent numerical simulations of magnetized stellar winds and specific dynamo and mass-loss prescriptions are adopted to tie angular momentum loss to angular velocity. The models additionally assume constant angular velocity during the disk accretion phase and allow for decoupling between the radiative core and the convective envelope as soon as the former develops.}{ We have developed rotational evolution models for slow, median, and fast rotators with initial periods of 10, 7, and 1.4d, respectively. The models reproduce reasonably well the rotational behavior of solar-type stars between 1~Myr and 4.5~Gyr, including pre-main sequence (PMS) to zero-age main sequence (ZAMS) spin up, prompt ZAMS spin down, and the early-main sequence (MS) convergence of surface rotation rates. We find the model parameters accounting for the slow and median rotators are very similar to each other, with a disk lifetime of 5~Myr and a core-envelope coupling timescale of 28-30~Myr. In contrast, fast rotators have both shorter disk lifetimes (2.5~Myr) and core-envelope coupling timescales (12~Myr). We show that a large amount of angular momentum is hidden in the radiative core for as long as 1~Gyr in these models and we discuss the implications for internal differential rotation and lithium depletion. We emphasize that these results are highly dependent on the adopted braking law. We also report a tentative correlation between the initial rotational period and disk lifetime, which suggests that protostellar spin down by massive disks in the embedded phase is at the origin of the initial dispersion of rotation rates in young stars.}{We conclude that this class of semi-empirical models successfully grasp the main trends of the rotational behavior of solar-type stars as they evolve and make specific predictions that may serve as a guide for further development. } 


   \maketitle
%

\section{Introduction}
The origin and evolution of stellar angular momentum still remains a mystery. Lately, a wealth of new observational constraints have been gained from the derivation of complete rotational distributions for thousands of low mass stars in young open clusters and in the field, covering an age range from 1~Myr to about 10~Gyr \citep[see, e.g.,][]{Irwin09a,Hartman10,Agueros11,Meibom2011,Irwin11,Affer2012,Affer2013}. These results now offer a detailed view of how surface rotational velocity changes as the stars evolve from the pre-main sequence (PMS), through the zero-age main sequence (ZAMS) to the late-main sequence (MS). A number of models have thus been proposed to account for these new observational results \citep[e.g.,][]{Irwin07,Bouvier08,Den10,Spada11,Reiners2012}. In order to satisfy observational constraints, most of these models have to incorporate three major physical processes: star-disk interaction during the PMS, angular momentum loss to stellar winds, and redistribution of angular momentum in the stellar interior. Indeed, each of these processes appears to have a fundamental role in dictating the evolution of surface rotation of solar-type stars from birth to the end of the main sequence and beyond. 

During the PMS, even though stars are contracting at a fast rate, they appear to be prevented from spinning up as long as they interact with their accretion disk, a process which lasts for a few Myr.  While the evidence for PMS rotational regulation is not recent \citep[see,][]{Edwards93,Bouvier93,Rebull04}, many theoretical advances have been made in the last years highlighting the impact of the accretion/ejection phenomenon on the angular momentum evolution of young suns \citep[e.g.,][]{Matt2012b,Zanni2012}. Similarly, it has long been known that low mass stars are braked on the MS as they lose angular momentum to their magnetized winds \citep{Sch62,Kraft67,WD67,Sku72,Kawaler88}. However, quantitative estimates of the angular momentum loss rates had to await the predictions of recent 2D and 3D numerical simulations of realistic magnetized stellar winds \citep[e.g.,][]{Vidotto2011,Aarnio12,Matt12}. Since observations have so far only revealed surface rotation, except for the Sun \citep[e.g.,][]{TC2011} and, more recently, for a few evolved giants \citep[e.g.,][]{Deheuvels12} thanks to asterosismology, the amount of angular momentum stored in the stellar interior throughout its evolution is usually unknown. Various mechanisms including hydrodynamical instabilities, internal magnetic fields, and gravity waves have been suggested that redistribute angular momentum from the core to the surface \citep[see, e.g.,][]{Spada10,Eggenberger12,Charbonnel13}. Obviously, as the star sheds angular momentum from the surface, the rate at which angular momentum is transported to the stellar interior has a strong impact on the evolution of surface rotation { \citep[e.g.,][]{Li93}}. 

The aim of the present study is to develop new angular momentum evolution models for solar-type stars, from 1~Myr to the age of the Sun, that incorporate some of the most recent advances described above and to compare their predictions to the full set of newly available observational constraints. One of the major differences between this study and previous similar studies lies in the wind braking relationship used in the models presented here that relies on recent stellar wind simulations by \citet{Matt12} and \citet{Cranmer11}.  In Sect.~\ref{sample}, we compile a set of 13 rotational period distributions that define the run of surface rotation as a function of age, which the models have to account for. In Sect.~\ref{modcons}, we describe the assumptions we used in the models to compute the angular momentum evolution of slow, median, and fast rotators, which include star-disk interaction, wind braking, and core-envelope decoupling.  The results are presented in Sect.~\ref{res} where the differences between fast and slow/median rotator models are highlighted. In Sect.~\ref{disc}, we discuss the implications of these models for internal differential rotation and disk lifetimes and provide a framework for understanding the evolution of the wide dispersion of rotational velocities observed for solar-type stars from the PMS to the early-MS. We conclude in Sect.~\ref{conc} by discussing the validity and limitations of these models

\section{Rotational distributions}
\label{sample}

\begin{table*}
\caption{Open clusters whose rotational distributions are used is this study.}             
\label{opencluster}      
\centering                          
\begin{tabular}{c c c c c c c c}       
\hline\hline                
Cluster & Age & N$_{star}$ & Mass bin & Ref. & $\Omega_{25}$ & $\Omega_{50}$ & $\Omega_{90}$ \\  
 &(Myr)& & ($\mathrm{M}_{\odot})$ &  & --- & ($\Omega_\odot$) & ---  \\
\hline   
ONC & 1 & 154 & 0.25-1.2 & 1 & {2.88$\pm$0.07} & {3.73$\pm$0.16} & {17.64$\pm$1.61} \\
NGC 6530 & 1.65 & 129 & 0.5-1.1 & 2 & {3.43$\pm$0.26} & {6.14$\pm$0.47} & {22.25$\pm$2.64} \\
NGC 2264 & 2 & 41 & 0.6-1.2 & 3 & {4.03$\pm$0.49} & {6.74$\pm$0.49} & {11.37$\pm$1.28} \\
NGC 2362 & 5 & 64 & 0.8-1.1 & 4 & {2.78$\pm$0.26} & {3.79$\pm$0.36} & {12.34$\pm$2.61} \\
h PER & 13 & 159 & 0.8-1.1 & 5 & {4.27$\pm$0.27} & {6.54$\pm$0.7} & {60.66$\pm$7.33} \\
NGC 2547 & 40 & 47 & 0.6-1.1 & 6 & {4.51$\pm$0.26} & {5.76$\pm$1.05} & {51.01$\pm$12.57} \\
Pleiades & 120 & 74 & 0.9-1.1 & 7 & {4.96$\pm$0.13} & {6.21$\pm$0.31} & {38$\pm$9.92} \\
M 50 & 130 & {62} & {0.9-1.1} & 8 & {3.81$\pm$0.31} & {5.21$\pm$0.33} & {36.34$\pm$19.02} \\
M 35 & 150 & {70} & {0.9-1.1} & 9 & {4.39$\pm$0.09} & {5.25$\pm$0.2} & {22.28$\pm$5.03} \\
M 37 & 550 & {75} & {0.9-1.1} & 10 & {2.95$\pm$0.05} & {3.26$\pm$0.06} & {3.75$\pm$0.16} \\
Praesepe & 578 & {12} & {0.9-1.1} & 11 & {2.54$\pm$0.05} & {2.68$\pm$0.04} & {2.73$\pm$0.03} \\
Hyades & 625 & {7} & {0.9-1.1} & 11 & {2.48$\pm$0.07} & {2.68$\pm$0.06} & {2.91$\pm$0.1} \\
NGC 6811 & 1000 & 31 & {0.9-1.1} & 12 & {2.25$\pm$0.02} & {2.33$\pm$0.02} & {2.39$\pm$0.02} \\
\hline   
\end{tabular}
\tablebib{(1)~\citet{Herbst02}; (2) \citet{Henderson11}; 
(3) \citet{Affer2013}; (4) \citet{Irwin08a}; (5) \citet{Moraux}; (6) \citet{Irwin08b}; (7) \citet{Hartman10}; (8) \citet{Irwin09b}; (9) \citet{Meibom09}; (10) \citet{Hartman09}; (11) \citet{Delorme11}; (12) \citet{Meibom2011}.}
\end{table*}

\begin{figure*}
\centering
\includegraphics[angle=-90,width=15cm]{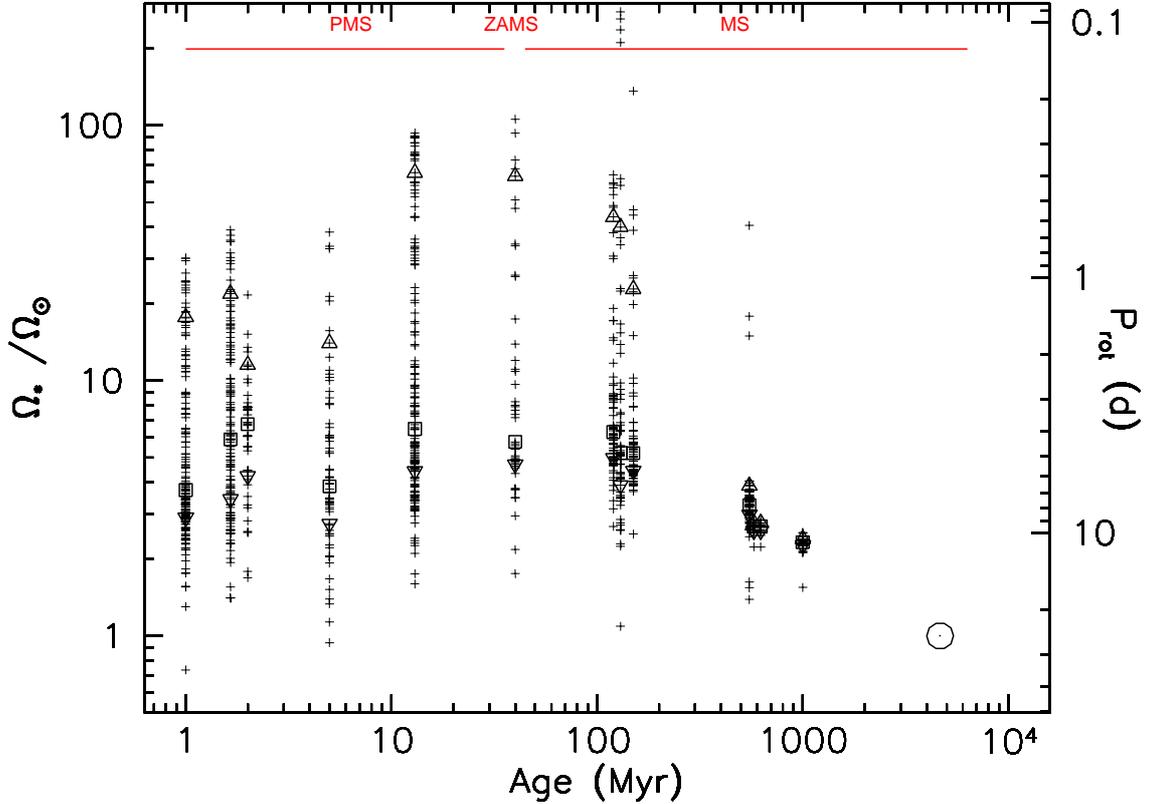}
\caption{Angular velocity distributions of solar-type stars in young open clusters and the Sun. Direct triangles, inverted triangles, and squares represent the $90^{th}$ percentile,  the $25^{th}$ percentiles, and the median of the observed distributions, respectively. The open circle shows the angular velocity of the present Sun. The left axis is labelled angular velocity normalized to the Sun's, while the right axis is labelled rotational periods (days).}
\label{amas}
\end{figure*}

In order to compare the angular momentum evolution models to observations, we used the rotational distributions measured for solar-type stars in 13 star forming regions and young open clusters covering the age range from 1~Myr to 1~Gyr, plus the Sun. The stellar clusters were chosen so as to provide rotational periods for at least 40 solar-type stars at a given age, a reasonable minimum for assessing the statistical significance of their distribution. However, this constraint was relaxed for older clusters, namely Praesepe (578~Myr), the Hyades (625~Myr), and NGC 6811 (1~Gyr), for which we used 12, 7, and 31 stars, respectively, as their rotational period distribution is single peaked and exhibits little dispersion \citep{Delorme11,Meibom2011}. Table \ref{opencluster} lists the 13 clusters used in this study. We originally selected  a stellar mass bin from  0.9 to 1.1~M$_\odot$ as being representative of the 1~M$_\odot$ rotational models. However, whenever no clear relationship existed between rotation and mass, we enlarged the mass bin to lower mass stars in order to increase the statistical significance of the rotational distributions. This is the case, for instance, for the Orion Nebula Cluster (ONC), where 154 stars with known rotational periods were selected over the mass range 0.25-1.2~M$_\odot$ (see Appendix \ref{clustparam} for details on the cluster parameters). 

All periods used here were derived by monitoring the rotational modulation of the stellar brightness due to surface spots (cf. references in Table~\ref{opencluster}). This method is free from inclination effects and provides a direct measurement of the star's rotational period, which is then easily converted to angular velocity ($\Omega_* = 2 \pi / P_{rot}$). The rotational distributions can, however, be affected by observational biases, such as rapid rotation in tidally synchronized binaries, slow rotation from contaminating field stars unrelated to the cluster, or aliases and/or harmonics of the true stellar period resulting from incomplete and/or uneven temporal sampling. We believe these biases do not strongly affect the percentiles of the distributions we use below to compare angular momentum evolution models to observations.

Figure \ref{amas} shows the angular velocity distribution of each cluster as a function of time, with angular velocities scaled to that of the Sun ($\Omega_\odot$=$2.87\times10^{-6}~\mathrm{s}^{-1}$). In this figure, the inverted triangles, squares, and direct triangles represent the $25^{th}$, $50^{th}$, and $90^{th}$ percentiles of the angular velocity distributions, respectively. The aim of the models presented below is to reproduce the run of these slow, median and fast rotators as a function of time. As these distributions suffer from statistical noise, we first estimate the error bars to be placed on the 3 percentiles. For each cluster, we applied a rejection method to randomly generate 5000 angular velocity distributions that follow the observed one. We then computed the percentiles of each of the synthetic distributions, thus yielding 5000 percentile estimates, $p_i$, at a given age. The median of the $p_i$ is taken as being the best estimate of the percentile value, and its associated error bar is computed as being the absolute deviation of individual estimates around the median, i.e., $\sigma = 1/n\sum_{i=1}^{n} |p_i - median(p_i)|$. The values of the percentiles and their error bars are listed in Table~\ref{opencluster}. The average difference between the computed median percentiles and the observed ones is 1.88 $\pm$ 1.83 $\%$ for slow rotators, 0.26 $\pm$ 1.34 $\%$ for median rotators, and 5.55 $\pm$ 5.98 $\%$ for fast rotators. Figure \ref{amas} shows that the rotational evolution of solar-type stars is now relatively well-constrained over the PMS and MS, with a nearly even sampling on a logarithmic age scale from 1~Myr to the age of the Sun. The models developed in this paper adopt the ONC rotational distribution as initial conditions at 1~Myr. This cluster exhibits a large spread in rotation rates, whose origin is currently unclear and points to the protostellar phase. We return to this point in Sect.~\ref{proto}.

\section{Model assumptions}
\label{modcons}

The angular momentum evolution of isolated solar-type stars depends mainly on three physical processes: angular momentum exchange within the evolving stellar interior, magnetic star-disk interaction in accreting young stars, and angular momentum removal by magnetized stellar winds. We discuss in turn the corresponding model assumptions in this section. 

\subsection{Internal structure}
In order to follow the evolution of the stellar structure, most notably during the pre-main sequence when the radiative core develops and the stellar radius shrinks, we adopt the \citet{Baraffe98} NextGen models computed for solar-mass stars of solar metallicity, with a mixing length parameter $\alpha = 1.5$, and helium abundance $Y = 0.275$. The model starts at 3 10$^3$ yr and yields $R_* = 1.02 R_{\odot}$ and $L_* = 1.04 L_{\odot}$ at  an age of 4.65 Gyr. Since the mass bins we selected in each cluster are representative of the rotation rates of solar-mass stars, we only used 1M$_\odot$ models. Furthermore, only solar metallicity models where used, neglecting the possible impact of a cluster's slightly different metallicity on the rotational properties of the members (cf. Appendix). Low mass stars are composed of two regions: an inner radiative core and an outer convective envelope. We follow \citet{McGB91} by assuming that both the core and the envelope rotate as solid bodies but with different angular velocity. The amount of angular momentum $\Delta J$  to be transferred from the core to the envelope in order to balance their angular velocities is given by 
\begin{eqnarray}
\Delta J = \frac{I_{env}J_{core}-I_{core}J_{env}}{I_{core}+I_{env}},
\end{eqnarray}
where $I$ and $J$ refer to the moment of inertia and angular momentum, respectively, of the radiative core and the convective envelope. As in \citet{Allain98}, we assume that $\Delta J$ is transferred over a time-scale $\tau_{c-e}$, which we refer to as the core-envelope coupling timescale. This is a free parameter of the model that characterizes the angular momentum exchange rate within the stellar interior. Previous modeling has shown that the coupling timescale may be different for fast and slow rotators \citep{Irwin07,Bouvier08,Irwin09a}, an issue to which we return below. {\citet{Den10} assumed a rotation-dependent coupling timescale and described it as a simple step function, where the coupling timescale is short at high velocity and suddenly increases below some critical velocity.} More recently, \citet{Spada11} have explored a {more} specific dependence of $\tau_{c-e}$ on rotation 
\begin{eqnarray}
\tau_{c-e} (t) = \tau_0 \left[\frac{\Delta \Omega_{\odot}}{\Delta \Omega (t)}\right]^\alpha,
\end{eqnarray}
where $\Delta \Omega_{\odot}$=0.2 $\Omega_{\odot}$, $\Delta \Omega (t)= \Omega_{core}-\Omega_{env}$, $\tau_0 = 57.7 \pm 5.24$~Myr,  and $\alpha = 0.076 \pm 0.02$. The derived dependency of $\tau_{c-e}$ on $\Delta\Omega$ is weak, and we simply assume here that $\tau_{c-e}$ is constant for a given model.

\subsection{Star-disk interaction}

For a few Myr during the early pre-main sequence, solar-type stars magnetically interact with their accretion disk, a process often referred to as magnetospheric accretion \citep[cf.][for a review]{Bouvier07b}. This star-disk magnetic coupling involves complex angular momentum exchange between the components of the system, including the accretion disk, the central star, and possibly both stellar and disk winds. Early models suggested that the magnetic link between the star and the disk beyond the corotation radius could result in a spin equilibrium for the central star \citep[e.g.,][]{Cameron93,Cameron95}. More recently, accretion-powered stellar winds have been proposed as a way to remove from the central star the excess of angular momentum gained from disk accretion \citep{MP05b,MP08a,MP08b}. However, \citet{Zanni11} showed that the characteristic accretion shock luminosity $\mathrm{L}_{UV}$ in young stars, of the order of $0.1~ \mathrm{L}_{\odot}$, implies that a significant fraction of the accretion energy is radiated through the accretion shock. They concluded that mass and energy supplied by accretion may not be sufficient to provide an efficient spin-down torque by accretion-driven winds. \citet{Zanni2012} propose that magnetospheric reconnection events occurring between the star and the disk lead to ejection episodes that remove the excess angular momentum. The issue of angular momentum exchange between the young star and its environment thus remains controversial and much work remains to be done to be able to provide a clear physical description of this process. Based on the observational evidence for a spin equilibrium in accreting young stars \citep{Bouvier93,Edwards93,Rebull04}, we simply assume here that the stellar angular velocity remains constant as long as the young star interacts with its disk. Hence, a free parameter of the models is the accretion disk lifetime $\tau_{disk}$, i.e., the duration over which the star's angular velocity is maintained at its initial value. After a time $\tau_{disk}$, the star is released from its disk, and is only subjected to angular momentum loss because as a result of magnetized stellar winds (see below). We note that during most of the pre-main sequence, once the disk has dissipated, angular momentum losses due to magnetized stellar winds are, however, unable to prevent the star from spinning up as its moment of inertia rapidly decreases towards the ZAMS \citep[cf.][]{Bouvier97,MP07}.

\subsection{Stellar winds}

Solar-type stars lose angular momentum as they evolve because of magnetized stellar winds \citep{Sch62,WD67}. 
Assuming a spherical outflow, the angular momentum loss rate due to stellar winds can be expressed as
\begin{eqnarray}
\label{djdt}
\frac{dJ}{dt} \propto \Omega_* \cdot \dot{M}_{wind} \cdot r_A^2,
\end{eqnarray}
where $r_A$ is the averaged value of the Alfv\'en radius that accounts for the magnetic lever arm, $\Omega_*$ is the angular velocity at the stellar surface, and $\dot{M}_{wind}$ is the mass outflow rate. Most angular momentum evolution models have so far used Kawaler's (1988) prescription to estimate the amount of angular momentum losses due to stellar winds, with some modifications such as magnetic saturation \citep{Kri97,Bouvier97} or a revised dynamo prescription \citep{Reiners2012}. The main difference between previous models and the ones we present here is that we base our estimates of angular momentum loss on the recent stellar wind simulations performed by \citet{Matt12} who derived the expression 
\begin{eqnarray}
\label{ranew}
r_A = K_1 \left[ \frac{B_p^2 R_*^2}{\dot{M}_{wind} \sqrt{K_2^2v_{esc}^2 + \Omega_*^2 R_*^2}}\right]^m R_*,
\end{eqnarray}
where $K_1 = 1.30$, $K_2 = 0.0506$, and $m = 0.2177$ are obtained from numerical simulations of a stellar wind flowing along the opened field lines of a dipolar magnetosphere. In Eq. \ref{ranew}, $R_*$ is the stellar radius, $B_p$ is the surface strength of the dipole magnetic field at the stellar equator, and $v_{esc}=\sqrt{2GM_*/R_*}$, where $M_*$ is the stellar mass, is the escape velocity. This equation is a modified version of the \citet{MP08a} prescription,
\begin{eqnarray}
\label{raold}
r_A = K_3 \left( \frac{B_p^2 R_*^2}{\dot{M}_{wind} v_{esc}} \right)^m R_*,
\end{eqnarray} 
where $K_3$ = 2.11 and m = 0.223 and are derived from numerical simulations. The difference between Eqs. \ref{ranew} and \ref{raold} is the term $\sqrt{1/\Omega_*^2 R_*^2}$  that takes into account how the Alfv\'enic radius depends on stellar rotation. 

In order to implement this angular momentum loss rate into our models, we have to express the Alfv\'enic radius as a function of stellar angular velocity only (and stellar parameters $M_*, R_*$). We must therefore adopt a dynamo prescription that relates the stellar magnetic field to stellar rotation, as well as a wind prescription that relates the mass-loss rate to the stellar angular velocity. We discuss now how to define such relationships,  based on theory and numerical simulations, and calibrated onto the present-day Sun.

\subsubsection{Dynamo prescription}
\label{magmass}

We assume the stellar magnetic field to be dynamo generated, i.e., that the mean surface magnetic field strength scales to some power of the angular velocity. We thus have 
\begin{eqnarray}
\label{b1}
 f_*B_* \propto \Omega_*^b,
\end{eqnarray}
where $b$ is the dynamo exponent, $B_*$ is the strength of the magnetic field, and $f_*$ is the filling factor, i.e., the fraction of the stellar surface that is magnetized \citep[cf.][]{Reiners2012}. Magnetic field measurements suggest that the magnetic field strength $B_*$ is proportional to the equipartition magnetic field strength $B_{eq}$ \citep[see][]{Cranmer11}
\begin{eqnarray}
B_* \approx 1.13~B_{eq},
\end{eqnarray}
where $B_{eq}$ is defined as
\begin{eqnarray}
B_{eq} = \sqrt{\frac{8 \pi \rho_* k_B T_{eff}}{\mu m_H}}
\end{eqnarray}
with $\rho_*$ the photospheric density, $k_B$ the Boltzmann's constant, $T_{eff}$ the effective temperature, $\mu$ the mean atomic weight, and $m_H$ the mass of a hydrogen atom. By using the magnetic field measurement of 29 stars, \citet{Cranmer11} found that the ratio $B_*/B_{eq}$ only slightly depends on the rotation period, i.e., $B_*/B_{eq} \propto P_{rot}^{-0.13}$, which implies that the magnetic field strength is almost constant regardless of the angular velocity. This behavior is consistent with the observations of \citet{Saar96} who found a slight increase of $B_*/B_{eq}$ for $P_{rot} < 3$ days \citep[see][Fig. 3]{Saar96}. In contrast, the magnetic filling factor $f_*$ appears to strongly depend on the Rossby number $Ro = P_{rot}/\tau_{conv}$, where $\tau_{conv}$ is the convective turnover time. According to \citet{Saar96} $f_* \propto P_{rot}^{-1.8}$, while \citet{Cranmer11} provide two different fits for $f_*$ that are, respectively, the lower and upper envelopes of the $f_*$-Ro plot (see their Fig. 7)
\begin{eqnarray}
f_{min} = \frac{0.5}{\left[1 + (x/0.16)^{2.6}\right]^{1.3}},
\end{eqnarray}
which is the magnetic filling factor linked to the open flux tubes in non-active magnetic regions, with $x = Ro/Ro_{\odot}$, $Ro_{\odot} = 1.96$, and 
\begin{eqnarray}
f_{max} = \frac{1}{1 + (x/0.31)^{2.5}},
\end{eqnarray}
which is linked to the closed flux tubes in active regions. Their empirical fits give $f_{min} \propto Ro^{-3.4}$ and $f_{max} \propto Ro^{-2.5}$, respectively. In the framework of our model the most relevant filling factor is $f_{min}$ which is related to the open flux tubes that carry matter through the stellar outflow. We therefore preferred the expression $f_{min}$, but slightly modified it in order to reproduce the average filling factor of the present Sun \citep[$f_{\odot}$ = 0.001-0.01, see Table 1 of][]{Cranmer11}
\begin{eqnarray}
f_* = \frac{0.55}{\left[1 + (x/0.16)^{2.3}\right]^{1.22}}.
\label{fmod}
\end{eqnarray}

We used the BOREAS\footnote{\url{https://www.cfa.harvard.edu/~scranmer/Data/Mdot2011/}} subroutine, developed by \citet{Cranmer11} to get the mean magnetic field $B_*f_*$ as a function of stellar density, effective temperature, and angular velocity. The photospheric density is calculated by BOREAS at the age steps provided by the \citet{Baraffe98} stellar structure models, and $f_*$ is derived form Eq. \ref{fmod} above. The upper panel of Fig. \ref{mdotbstarjdot} shows the resulting mean magnetic field strength as a function of stellar angular velocity. It is seen that $B_*f_*$ increases from $\Omega_{*} \simeq 1~\Omega_{\odot}$ to $\Omega_{*} \simeq 10~\Omega_{\odot}$, and then starts to saturate at $\Omega_{*} \geq 15~\Omega_{\odot}$. We derive the following asymptotic expressions for the slow and fast rotation regimes, respectively,
\begin{flalign}
\label{Bnonsat}
& f_*B_* (G)  \simeq 7.3~\left(\frac{\Omega_*}{\Omega_{\odot}}\right)^{2.6}& &\text{if $1.5~\Omega_{\odot} \leq \Omega_{*} \leq 4~\Omega_{\odot}$,}  \\
\label{Bsat}
& f_*B_* (G) \simeq 910& &\text{if $\Omega_{*} \geq \Omega_{sat}$,} 
\end{flalign}
where $\Omega_{sat} \approx 15 ~\Omega_{\odot}$. The saturation threshold is dictated by the expression of $f_*$ that we adopt in our simulation. We used the Rossby prescription from \citet{Cranmer11}, i.e., for a solar-mass star $\tau_{conv} \approx 30$d at 10 Myr,  decreasing to 15d at an age $\geq 30$ Myr. 
Measurements of stellar magnetic fields suggest that saturation is reached at $Ro \lesssim 0.1-0.13$ \citep[see][Fig. 6]{Reiners09}. With $\tau_{conv} \approx$ 15~days, this translates into a dynamo saturation occuring at $\Omega_{sat} \sim 13-17~\Omega_{\odot}$, which is consistent with the value we derive here (cf. Fig.~\ref{mdotbstarjdot}). In Eq. \ref{ranew}, $B_p$ is the strength of the dipole magnetic field at the stellar equator. Even though the real stellar magnetic field is certainly not a perfect dipole, we identify $B_p$ to the strength of the mean magnetic field $B_*f_*$.

\subsubsection{Wind loss rate prescription}
\label{windpresc}

The mass-loss rate of solar-type stars at various stages of evolution is unfortunately difficult to estimate directly from observation. We therefore have to rely mainly on the results of numerical simulations of stellar winds, calibrated onto a few, mostly indirect, mass-loss measurements \citep[e.g.,][]{Wood02,Wood05a}. Here we used the results from the numerical simulations of \citet{Cranmer11}. Assuming that the wind is driven by gas pressure in a hot corona, as is likely the case for G-K stars, they found $\dot{M}_{wind} \propto f_*^{5/7}$. As we did in the case of the mean magnetic field, we used the output of the BOREAS subroutine to get the mass-loss rate as a function of several stellar parameters such as the angular velocity, the luminosity, and the radius. In particular, the mass-loss rate strongly depends on the quantity of energy $F_{A*}$ deposited by the Alfvén waves as they propagate through the photosphere and are subsequently converted into a heating energy flux that powers the stellar wind \citep[see][for details]{Musielak1,Musielak3,Musielak2}. \citet{Cranmer11} provided an analytical fit, based on the results of \citet{Musielak2}, for $F_{A*}$ in the case where the mixing length parameter $\alpha = 2$ and $B_* / B_{eq} = 0.85$. In our model, we use $\alpha = 1.5$ and $B_* / B_{eq} = 1.13$, and estimate that \citet{Cranmer11} overestimates $F_{A*}$ by a factor of about 5 \citep[see Eqs. 14 and 15 from][]{Musielak2}. We empirically adopt a dividing factor of 2.5 in our model to recover a mass-loss rate of $1.42 \times 10^{12}$g.s$^{-1}$ at the age of the Sun, which is consistent with the estimated range of the present Sun's mass-loss rate \citep[$\dot{M}_{\odot} = 1.25-1.99 \times 10^{12}$g.s$^{-1}$, see Table 2 from][]{Cranmer11}. 
\begin{figure}
\centering
\includegraphics[angle=-90,width=9cm]{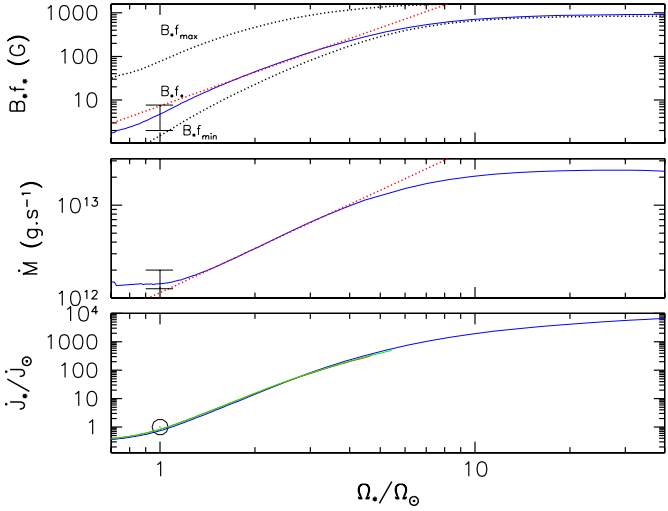}
\caption{\textit{Upper panel:} Mean magnetic field strength computed from the BOREAS subroutine as a function of stellar angular velocity normalized to the Sun's velocity. The Sun's range of $B_*f_* = 2-7.7~G$ is shown as a vertical bar. The upper and lower dotted lines illustrate $B_*f_{max}$ and $B_*f_{min}$, respectively. The red dashed line is a power-law fit to $B_*f_*$ in the non-saturated regime (cf. Eq. \ref{Bnonsat}). \textit{Middle panel:} The mass-loss rate computed from the BOREAS subroutine as a function of stellar angular velocity normalized to the Sun's velocity. The range of $\dot{M}$ estimate for the Sun is shown with a vertical bar. The red dashed line is a power-law fit to $\dot{M}$ in the non-saturated regime (cf. Eq. \ref{mdot1}). \textit{Lower panel:} The angular momentum loss rate as a function of angular velocity. Both quantities are normalized to the Sun's ($\dot{J}_{\odot} = 7.169 \times 10^{30} ~ \mathrm{g.cm^2.s^{-2}}$, and $\Omega_{\odot} = 2.87 \times 10^{-6} ~ \mathrm{s^{-1}}$). Three overlapping curves are illustrated for fast (blue), median (green), and slow (red) rotator models. The temporal evolution of $T_{eff}$ and $R_*$ is included in the computation of $\dot{J}/\dot{J_\odot}$ for each model.}
\label{mdotbstarjdot}
\end{figure}
  
The middle panel of Fig. \ref{mdotbstarjdot} shows the evolution of the mass-loss rate as a function of stellar angular velocity in our models.  The saturation of the mass-loss rate again appears around 10 $\Omega_{\odot}$,  corresponding to the saturation of $f_*$. We derive the following asymptotic expressions for the mass loss-rate prescription in the slow and fast rotation regimes, respectively,
\begin{flalign}
\label{mdot1}
&\dot{M}_{wind} \simeq 1.14\times10^{12} \left(\frac{\Omega_*}{\Omega_{\odot}}\right)^{1.58}~g.s^{-1}, &
\end{flalign}
if $1.5~\Omega_{\odot} \leq \Omega_{*} \leq 4~\Omega_{\odot}$, and
\begin{flalign}
\label{mdot2}
&\dot{M}_{wind} \simeq 2.4\times10^{13}~g.s^{-1}, &
\end{flalign}
if $\Omega_{*} \geq \Omega_{sat}$, where $\Omega_{sat} \approx 15 ~\Omega_{\odot}$.

\subsubsection{Angular momentum loss rate: asymptotic forms} 

To highlight the dependency of the angular momentum loss rate on stellar parameters and primarily on stellar angular velocity, we express $dJ/dt$, in the asymptotic cases of slow and fast rotators, as a power law  combining Eqs. \ref{djdt}, \ref{ranew}, \ref{Bnonsat}, \ref{Bsat}, \ref{mdot1}, and \ref{mdot2} above, to yield
\begin{eqnarray}
\label{djdtmodel12}
\frac{dJ}{dt} &=& 1.22\times10^{36} \frac{K_1^2 R_*^{3.1}}{\left(K_2^22GM_*+\Omega_*^2R_*^3\right)^{0.22}} \Omega_{*}^{4.17}
\end{eqnarray}
\noindent if $1.5~\Omega_{\odot} \leq \Omega_{*} \leq 4~\Omega_{\odot}$, and
\begin{eqnarray}
\label{djdtmodel22}
\frac{dJ}{dt} &=& 2.18\times10^{16} \frac{K_1^2 R_*^{3.1}}{\left(K_2^22GM_*+\Omega_*^2R_*^3\right)^{0.22}} \Omega_{*}
\end{eqnarray}
in the saturated regime ($\Omega_* \geq 15~\Omega_{\odot}$). Fig \ref{mdotbstarjdot} shows how the angular momentum loss rate varies with angular velocity for the three rotational models developed below. 

\section{Results}
\label{res}

\begin{figure*}
\centering
\includegraphics[angle=-90,width=15cm]{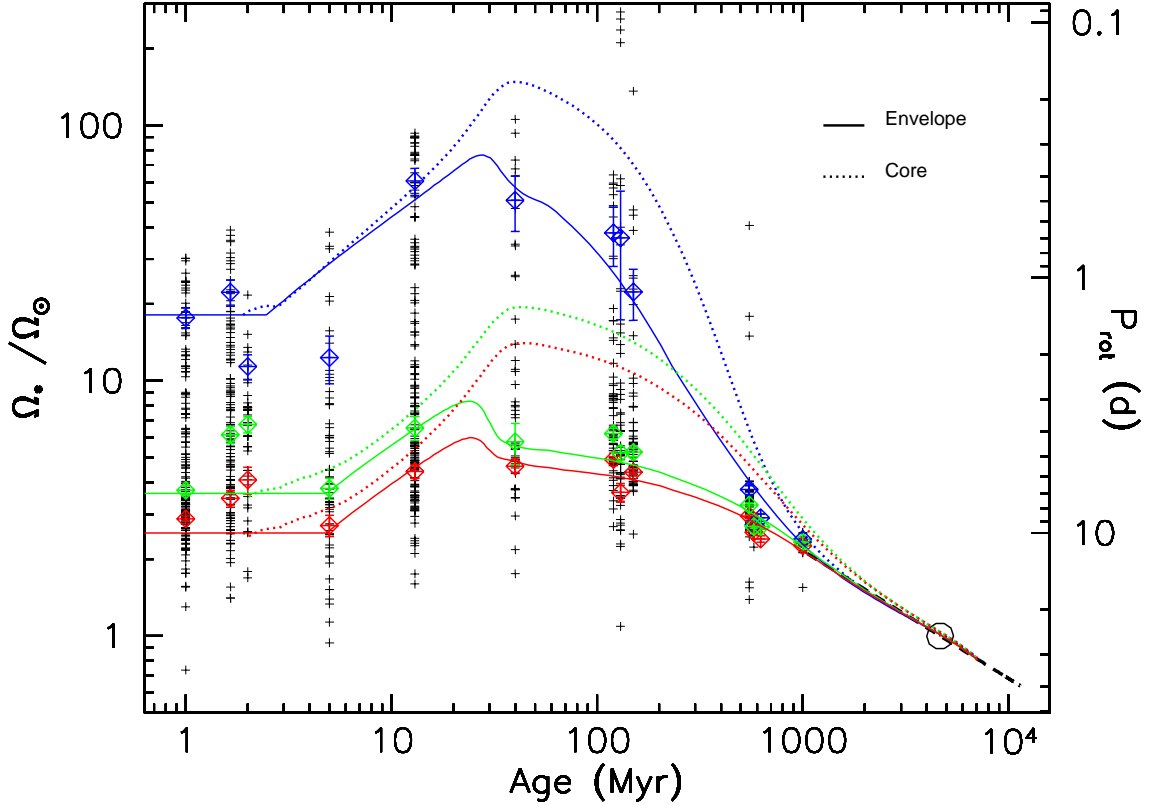}
\caption{Angular velocity of the radiative core (dashed lines) and of the convective envelope (solid lines) is shown as a function of time for fast (blue), median (green), and slow (red) rotator models. The angular velocity is scaled to the angular velocity of the present Sun. The blue, red, and green tilted squares and associated error bars represent the $90^{th}$ percentile, the $25^{th}$ percentile, and the median, respectively, of the rotational distributions of solar-type stars in star forming regions and young open clusters obtained with the rejection sampling method (see text). The open circle is the angular velocity of the present Sun and the dashed black line illustrates the Skumanich relationship, $\Omega\propto t^{-1/2}$.}
\label{model}
\end{figure*}
 
The free parameters of the model are the initial rotational period at 1~Myr $P_{init}$, the core-envelope coupling timescale $\tau_{c-e}$, the disk lifetime $\tau_{disk}$, and the scaling constant of the wind braking law $K_1$. The value of these parameters are to be derived by comparing the models to the observed rotational evolution of solar-type stars. The models for slow, median, and fast rotators are illustrated in Fig. \ref{model} and their respective parameters are listed in Table~\ref{param}. As explained below, the initial period for each model is dictated by the rotational distributions of the youngest clusters, while the disk lifetime is adjusted to reproduce the observed spin up to the 13~Myr h~Per cluster. We did not attempt any chi-square fitting but merely tried to reproduce by eye the run of the rotational percentiles as a function of time.

\begin{figure} 
\centering 
\includegraphics[angle=-90,width=9cm]{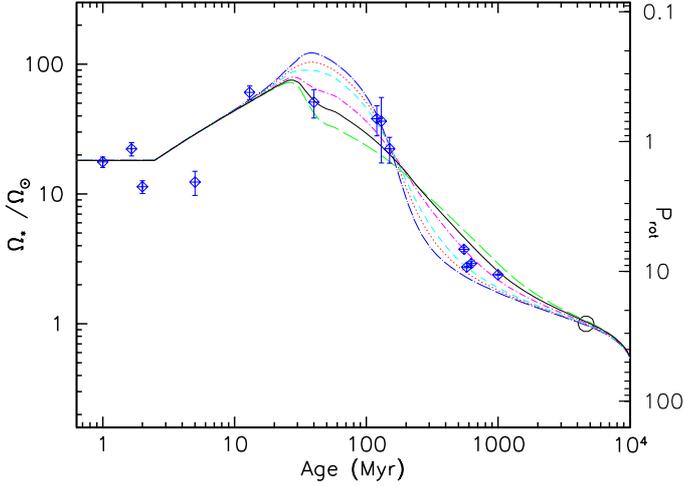} 
\caption{Angular velocity evolution for different values of the coupling time-scale $\tau_{c-e}$ for the fast rotator model ($P_{init} = 1.4$ days, $\tau_{disk}=2.5$ Myr). From top to bottom at the ZAMS the values for $\tau_{c-e}$ are: $1$ Myr (blue dot - long-dashed line), $3$ Myr (red dotted line), $5$ Myr (cyan short-dashed line), $10$ Myr (magenta dot - short-dashed line), $15$ Myr (black solid line), and $20$ Myr (green long-dashed line). The blue tilted square and associated error bars represent the $90^{th}$ percentile of the rotational distributions of solar-type stars in star forming regions and young open clusters obtained with the rejection sampling method (see text). The open circle is the angular velocity of the present Sun.}
\label{taudec}
\end{figure}

\begin{figure}
\centering
\includegraphics[angle=-90,width=9cm]{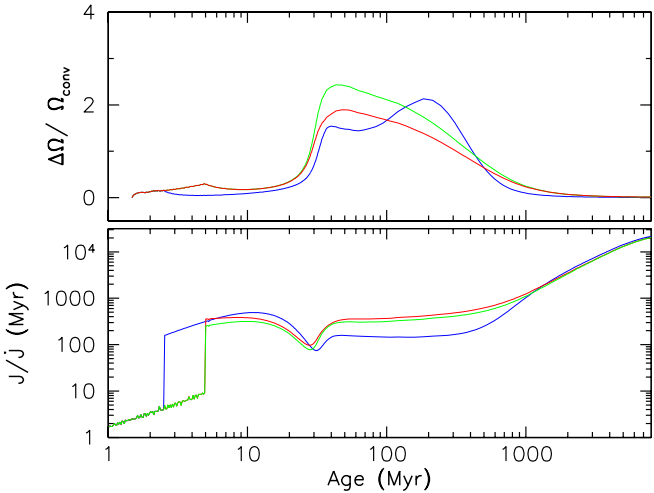}
\caption{\textit{Upper panel:} Velocity shear at the base of the convective zone $(\Omega_{core}-\Omega_{env})/\Omega_{env}$ in the case of fast (blue), median (green), and slow (red) rotator models. \textit{Lower panel:} Spin-down time-scale ($J/\dot{J}$) expressed in Myr, in the case of fast (blue), median (green), and slow (red) rotator models.}
\label{sddww}
\end{figure}

For the fast rotator model ($P_{init} = 1.4d$), the disk lifetime is taken to be as short as 2.5 Myr, resulting in a strong PMS spin up. This is required to fit the rapid increase of angular velocity between the youngest clusters at a few Myr ($\Omega_*\simeq 10-20~\Omega_\odot$) and the 13 Myr h Per Cluster ($\Omega_*\simeq 60~\Omega_\odot$). The choice of $P_{init} = 1.4$~d for this model is dictated by the fast rotators in the two youngest clusters (ONC and NGC 6530). However, it is seen from Fig. \ref{model} and Table~\ref{opencluster} that slightly older PMS clusters (NGC 2264 and NGC 2362) do not appear to harbor such fast rotators. Whether this is due to statistical noise or observational biases, or whether it actually reflects different cluster-to-cluster initial conditions, possibly linked to environmental effects \citep[cf.][]{Littlefair10,Bolmont2012}, is yet unclear. The core-envelope coupling timescale of the fast rotator model is 12 Myr which is comparable to the 10 Myr coupling timescale adopted by \citet{Bouvier08}, but much longer than the 1~Myr value used in \citet{Den10}. The reason for this difference is twofold.  First, the adoption of different braking laws results in different coupling timescales that reproduce the same set of rotational distributions. Second, the inclusion in our work of the recently derived h~Per rotational distribution at 13~Myr \citep{Moraux} yields new constraints on pre-ZAMS spin up that were not accounted for in previous studies. Figure~\ref{taudec} clearly shows that a coupling timescale as short as 1~Myr would not fit the observed evolution of fast rotators around the ZAMS.  Still, the 12~Myr coupling timescale we derive is short enough to allow the core and the envelope to exchange a large amount of angular momentum. In this way, the whole star is accelerated and the envelope reaches the high velocities observed at the ZAMS ($\Omega_*\simeq 50-60~\Omega_\odot$). A longer coupling timescale would fail to account for the fastest rotators on the ZAMS. This is clearly shown in Fig. \ref{taudec} that illustrates the impact of the coupling timescale $\tau_{c-e}$ on the rotational evolution of the envelope in the fast rotator model. Longer coupling timescales yield lower rotation rates on the ZAMS, as the inner radiative core retains most of the angular momentum while the convective envelope starts to be spun down. While for $\tau_{c-e}$ = 15~Myr, the velocity at ZAMS reaches $50~\Omega_{\odot}$, for $\tau_{c-e}$ = 1~Myr it amounts to $120~\Omega_{\odot}$. So, a relatively short coupling timescale of 10-15 Myr is required in order to fit the observational constraints, i.e., initial conditions, fast PMS spin-up, and high rotation rates on the ZAMS, from the PMS to the ZAMS. The choice of the coupling timescale also has an impact the shape of the angular velocity evolution on the early MS (Fig. \ref{taudec}). A short coupling timescale leads to a steeper spin down on the early MS, as the fastest ZAMS rotators are more efficiently braked by stellar winds. For longer coupling timescales, the early MS spin down is shallower, which arises from both a weaker angular momentum loss at the stellar surface and the angular momentum stored in the core being transferred back to the envelope on a timescale of $\simeq$100~Myr. The comparison of the models with the observations suggests that a core-envelope coupling timescale of 10-15~Myr best reproduces the spin-down rate of fast rotators on the early MS. In these models, the largest amount of differential rotation between the inner radiative core and the outer convective envelope is reached at 200 Myr and amounts to $\Delta\Omega/\Omega\simeq 2-2.5$ (cf. Fig.~\ref{sddww}).

As can be seen from Table \ref{param}, the parameters for the median and slow rotator models are quite similar to each other. The initial rotational periods are 7d and 10d for the median and slow rotator models, respectively, as indicated by the rotational distributions of the youngest PMS clusters, with significant scatter, however, over the first 5 Myr (see above). For both models we chose a disk lifetime of 5~Myr in order to reproduce the late PMS clusters and the slow rotation rates still observed in the 13~Myr h~Per cluster ($\Omega_*\leq 7~\Omega_\odot$).  To account for the weak PMS spin up of the envelope, which leads to moderate velocities on the ZAMS ($\Omega_*\leq 6~\Omega_\odot$), we had to assume a much longer core-envelope coupling timescale than for fast rotators, namely 28 and 30~Myr for median and slow rotator models, respectively. These values are significantly smaller than the 100 Myr coupling timescale derived by \citet{Bouvier08} and comparable to the value of 55$\pm$25~Myr derived by \citet{Den10}. The longer coupling timescale \citet{Bouvier08} assumes for slow rotators stems from the Kawaler braking law used in those models, which predicts weaker spin-down rate for slow rotators than the braking law we adopt here. Indeed, the slow rotation rates observed at 40~Myr requires the convective envelope to be braked {\it before} the star reaches the ZAMS, which suggests that only the outer convective envelope is spun down while the inner radiative core continues to accelerate all the way to the ZAMS (cf. Fig. \ref{model}). These models thus suggest that strong differential rotation develops between the radiative core and the convective envelope, reaching a maximum value of $\Delta\Omega/\Omega\simeq 2-2.5$  at 40~Myr, i.e., at the start of the MS evolution (cf. Fig.~\ref{sddww}). A long coupling timescale also implies a long-term transfer of angular momentum from the core to the envelope, which nearly compensates for the weak angular momentum loss at the stellar surface on the early MS. Thus, in sharp contrast to the fast rotator models which predict a factor of 10 decrease in rotation rate from the ZAMS to the Hyades age, the models for slow and median rotators predict a much shallower decline of surface rotation on the early MS, amounting to merely a factor of $\simeq$ 2 over the age range 0.1-1.0~Gyr. The slow decline of surface rotation is due to angular momentum of the core resurfacing at the stellar surface on a timescale of $\simeq$100~Myr in slow and moderate rotators and it accounts for the observed evolution of the lower envelope of the rotational distributions of early MS clusters (cf. Fig. \ref{model}). 

By 1~Gyr, the slow, median, and fast rotators models have all converged  towards the same surface angular velocity and stars are thereafter braked at a low pace, following Skumanich's relationship \citep{Sku72}, i.e., $\Omega_*\propto t^{-1/2}$. It is quite noticeable, however, that this relationship is not valid earlier on the MS, nor does a unique relationship between age and surface rotation prior to about 1~Gyr for solar-type stars \citep{Epstein2012}. All the models presented here yield a complete recoupling between the radiative core and the convective envelope by the age of the Sun, as requested by helioseismology results \citep{Thom03}. We emphasize that the evolution of core rotation strongly depends on the core-envelope coupling timescale assumed in the models and currently lacks observational constraints, apart from the solar case.
\begin{table}
\begin{center}
\caption{Model parameters.} 
\label{param}          
\begin{tabular}{c c c c}     
\hline\hline              
Parameter & Slow & Median & Fast \\  
\hline
$P_{init}$ (days) & 10 & 7 & 1.4 \\
$\tau_{c-e}$ (Myr) & 30 & 28 & 12 \\
$\tau_{disk}$ (Myr) & 5 & 5 & 2.5 \\
$K_1$ & 1.8 & 1.8 & 1.7 \\
\hline                                  
\end{tabular}
\end{center}
\end{table}

\begin{figure}
\centering
\includegraphics[angle=-90,width=9cm]{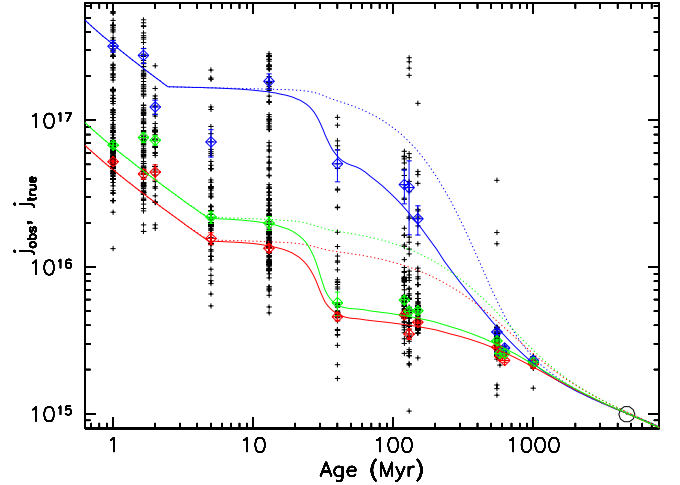}
\caption{Observed specific angular momentum ($j_{obs}=I_* \Omega_{env}/M_{*}$) evolution (solid line) and actual specific angular momentum ($j_{true}=(I_{core} \Omega_{core}+I_{env} \Omega_{env})/M_{*}$) evolution (dotted line) for the fast (blue), median (green), and slow (red) rotator models. The blue, green, and red tilted squares represent respectively the $90^{th}$, $50^{th}$, and $25^{th}$ percentiles of the observed specific angular momentum computed from the cluster's rotational distributions. The open circle is the specific angular momentum of the present-day Sun.}
\label{amspec}
\end{figure}

In Fig. \ref{amspec}, we show the same models for slow, median, and fast rotators where angular velocity has been converted to specific angular momentum. We define the observed specific angular momentum as $j_{obs}=I_* \Omega_{env}/M_{*}$, which would be derived from the surface angular velocity by assuming the star is a solid-body rotator (i.e., $\Omega_{core}=\Omega_{env}$), and the actual specific angular momentum $j_{true}=(I_{core} \Omega_{core}+I_{env} \Omega_{env})/M_{*}$, which takes into account the different rotation rates between the core and the envelope, as predicted by the models. We used the \citet{Baraffe98} evolutionary models to estimate $I_{core}$, $I_{conv}$, and $I_*$, for solar-mass stars at the ages of each cluster. Their respective values, normalized to $I_{\odot}=J_{\odot}/\Omega_{\odot}=6.41\times 10^{53}$~g.cm$^2$, are listed in Table~\ref{interpol}. During the early PMS, as long as the nearly fully convective star is coupled to the disk, the assumption of constant angular velocity translates into a significant decrease of specific angular momentum as the stellar radius shrinks ($j\propto \Omega_* R_*^2$). Because the star is released from the disk at a few Myr, angular momentum losses due to stellar winds are weak, and $j$ does not vary much for the next 10-20~Myr. Closer to the ZAMS, however, because fast rotators have reached their maximum velocity and slow rotators have experienced core-envelope decoupling, $j_{obs}$ will decrease again. In slow rotators, most of the angular momentum remains hidden in the inner radiative core. The different evolution of $j_{obs}$ and $j_{true}$ seen in Fig. \ref{amspec} past the ZAMS clearly illustrates the storage of angular momentum in the radiative core that is gradually transferred back to the convective envelope on a timescale of several 100~Myr. Eventually, all the models converge to the specific angular momentum of the present-day Sun by 4.56~Gyr \citep[$j_{\odot}\approx 9.25\times10^{14}~\mathrm{cm}^2.\mathrm{s}^{-1}$;][]{Pinto11}. 
\begin{table}
\begin{center}
\caption{Radius and moment of inertia of solar-mass stars.}           
\label{interpol}      
\begin{tabular}{c c c c c c} 
\hline\hline                
Cluster & Age & Radius & $I_{env}$ & $I_{core}$ & $I_{star}$\\
& (Myr) & ($R_{\odot}$)  & --- & ($I_{\odot}$) & ---\\
\hline                       
ONC & 1 & 2.58 & 19.6 & 0 & 19.6 \\ 
NGC 6530 & 1.65 & 2.12 & 13.4 & 0 & 13.4 \\ 
NGC 2264 & 2 & 1.98 & 11.7 & 0.004 & 11.7 \\ 
NGC 2362 & 5 & 1.44 & 5.7 & 0.5 & 6.2 \\
h PER & 13 & 1.12 & 2.1 & 1.2 & 3.3 \\ 
NGC 2547 & 40 & 0.92 & 0.13 & 0.93 & 1.06 \\ 
Pleiades & 120 & 0.9 & 0.13 & 0.9 & 1.03 \\
M50 & 130 & 0.9 & 0.13 & 0.9 & 1.03 \\ 
M35 & 150 & 0.9 & 0.13 & 0.9 & 1.03 \\ 
M37 & 550 & 0.91 & 0.12 & 0.91 & 1.03 \\
Praesepe & 578 & 0.91 & 0.12 & 0.91 & 1.03 \\ 
Hyades & 625 & 0.91 & 0.12 & 0.91 & 1.03 \\ 
NGC 6811 & 1000 & 0.92 & 0.12 & 0.95 & 1.07 \\
\hline   
\end{tabular}
\end{center}
\end{table}

\begin{figure}
\centering
\includegraphics[angle=-90,width=9cm]{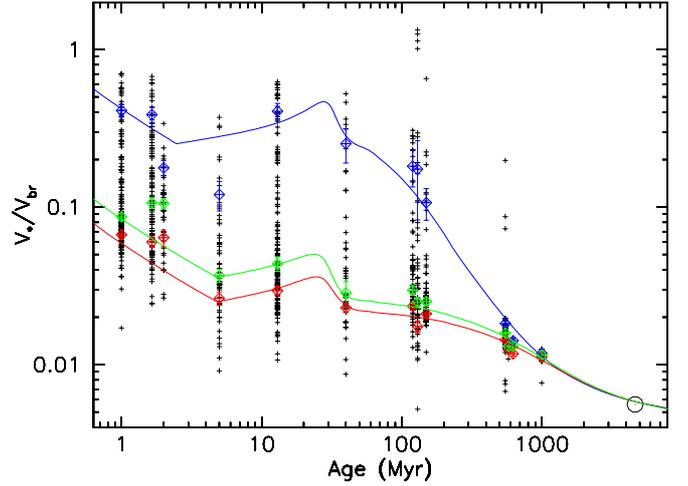}
\caption{Evolution of surface velocity scaled to break-up velocity for fast (blue), median (green), and slow (red) rotator models. The blue, red, and green tilted squares represent respectively the $90^{th}$ quartile, the $25^{th}$ quartile, and the median of the rotational distributions. The open circle represents the Sun. }
\label{breakup}
\end{figure}

Finally, both models and observations are shown in Fig. \ref{breakup} where the surface velocity has been normalized to the break-up velocity $V_{br} =\left( \frac{2}{3} \right)^{1/2} \sqrt{GM_*/R_*}$ where the factor $\left( \frac{2}{3} \right)^{1/2}$ comes from the ratio of the equatorial to the polar radius at critical velocity. The radius of non-rotating solar-mass stars at the age of the various clusters has been obtained from the \citet{Baraffe98} evolutionary models (see Table \ref{interpol}). As the stellar radius shrinks during the PMS, the break-up velocity of a solar-mass star increases from $ 222$ km/s at 1~Myr to $371$ km/s at the ZAMS. As long as the star is coupled to the disk, $\Omega_*$ is held constant, i.e., $V_{*} \propto R_*$ while $V_{br}$ increases, resulting in a net decrease of $V_*/V_{br} \propto R^{3/2}$ (Fig. \ref{breakup}). At $t = \tau_{disk}$, the star begins to spin up as it contracts towards the ZAMS at a faster rate ($V_*\propto R^{-1}$) than the increase of the break-up velocity, which results in the increasing $V_*/V_{br} \propto R^{-1/2}$ seen in Fig. \ref{breakup} prior to the ZAMS. At this stage, fast rotators can reach about 40-50\% of the break-up velocity. Later on the MS, the velocity of the fast rotators decreases from about $0.15~V_{br}$ at 100~Myr to $10^{-2}~V_{br}$ at 1~Gyr. The median/slow rotator models start at 0.08 and 0.05 $V_{br}$, respectively, at the age of the ONC. The angular velocity predicted by these models never exceeds 0.06 $V_{br}$ from the ZAMS to the age of the Sun. All models eventually reach $V_* \simeq 10^{-2}~V_{br}$ at $\approx$ 1~Gyr. We note that the few outliers with velocities close to and beyond the break-up velocity at an age of 130~Myr in Fig.~\ref{breakup} are probably either field contaminants unrelated to the M50 cluster, or contact binaries whose rotational evolution is driven by tidal effects \citep{Irwin09b}. 

\section{Discussion}
\label{disc}

The evolution of surface rotation of solar-type stars is now well documented, from their first appearance in the HR diagram as $\simeq$1~Myr PMS stars up to the age of the Sun, thanks to the measurement of thousands of rotational periods in star forming regions and young open clusters. Three main phases can be identified: a nearly constant surface rotation rate for the first few million years of PMS evolution, a rapid increase during the late PMS up to the ZAMS, followed by a slower decline on a timescale of a few 100~Myr on the early MS. In addition, observations indicate an initially wide dispersion of rotation rates at the start of the PMS, with a range of rotational periods extending from 1-3 days to 8-10 days. The initial dispersion increases further on the ZAMS, with periods ranging from 0.2-0.4 days to about 6-8 days, then subsequently decreases along the MS as surface rotation eventually converges to periods of the order of 10-12 days around 1~Gyr. Angular momentum evolution models aim at reproducing both the observed run of surface rotation with time and the evolution of the rotational dispersion as the stars age. Building up on previous modeling efforts, the models presented here suggest that these trends can be successfully reproduced with a small number of assumptions: i) a magnetic star disk interaction during the early PMS that prevents the young star from spinning up, thus accounting for a phase of nearly constant surface rotation as long as the star accretes from its disk; ii) angular momentum loss from magnetized stellar winds, a process that is instrumental as soon as the disk disappears but whose effects start to be felt only when the stellar contraction is nearing completion at the end of the PMS, thus still allowing PMS spin up before MS braking takes over; and iii) redistribution of angular momentum in the stellar interior, which allows part of the initial angular momentum to be temporarily stored in the inner radiative core while the outer convective envelope is spun down on the MS. The combination of star-disk interaction, wind braking, and core-envelope decoupling thus fully dictates the surface evolution of solar-type stars. We discuss in the following sections the impact of each of these physical processes on the models.

\subsection{Core-envelope decoupling and the shape of the gyrotracks}
\label{cedec}

\begin{table*}
\begin{center}
\caption{Comparison of the wind braking prescriptions.}   
\label{compbrake}    
\begin{tabular}{c c c c}        
\hline\hline                
Models& \textit{This study} & \textit{Reiners \& Mohanty 2012} & \textit{Bouvier 2008}\\
Braking law & \textit{Matt et al. 2012} & \textit{Reiners \& Mohanty 2012} & \textit{Kawaler 1988} \\
\hline                       
Solar calibration ($\dot{J}(\Omega=\Omega_\odot))$ & $4.6\times 10^{30}$ & ${4.1\times10^{31}}$ & ${8.9\times 10^{30}}^a$ \\
&&&${1.8\times 10^{31}}^b$ \\
$\dot{M}$ ($M_{\odot}/\mathrm{yr}$) & $\simeq 1.8\times 10^{-14}\times {\Omega_*^{1.58}/\Omega_\odot}^b$ & $1\times10^{-10}$ & No mass loss \\
$\Omega_{sat,B}$ & $\simeq$ 15 $\Omega_{\odot}$ & 3 $\Omega_{\odot}$ & 8 $\Omega_{\odot}$ \\
$\Omega_{sat,\dot{M}}$ & $\simeq$ 15 $\Omega_{\odot}$ & No sat. & N/A\\
$Asymptotic~\dot{J}^b$ & $\propto \Omega_*^{4.17}$ & $\propto \Omega_*^{5}$ & $\propto \Omega_*^{3}$ \\
$Asymptotic~\dot{J}^a$ & $\propto \Omega_*^{0.56}$ & $\propto \Omega_*$ & $\propto \Omega_*$ \\
\hline    
\end{tabular}
\end{center}
\textit{a}: fast rotators; \textit{b}: slow rotators 
\end{table*}

The balance between wind braking and internal angular momentum redistribution dictates the {\it shape} of the rotational tracks (hereafter called gyrotracks) which may vary between slow and fast rotators. For cases of strong core-envelope coupling, the whole star reacts to angular momentum loss at the stellar surface, which results in a long-term, steady decline of the surface velocity. On the contrary, for largely decoupled models, the radiative core retains most of the initial angular momentum and the outer convective envelope is rapidly braked owing to its reduced moment of inertia and, at later times, the angular momentum stored in the radiative core resurfaces into the envelope, thus delaying the spin-down phase. The core-envelope decoupling assumption used here yields a discontinuity of the angular velocity at the core-envelope interface, and should be considered as a simple-minded approximation of more physically-driven internal rotational profiles \citep[e.g.,][]{Spada10,Den10,Brun11,TC2011,Lagarde12}. However, regardless of the actual rotational profile solar-type stars develop as they evolve, the important point here is that models do allow angular momentum to be hidden in the inner region of the star, which subsequently resurfaces on evolutionary timescales. This is the key of the differences exhibited by the slow and fast rotator models presented in the previous section. Fast rotators have relatively short core-envelope coupling timescale, of the order of 12~Myr, which ensures both efficient PMS spin up in order to reach equatorial velocities up to $\simeq$ 80-125~km/s at ZAMS {\it and} a steady, monotonic spin down on the MS down to velocities of $\simeq$ 4.2~km/s at 1~Gyr. The slow and median rotators ($V_{eq}\leq$10-16~km/s) on the other hand have longer coupling timescales of the order of 28-30~Myr. This allows the envelope to be efficiently braked before the star reaches the ZAMS in spite of overall PMS spin up, thus explaining the significant number of slow rotators on the ZAMS, and simultaneously accounts for their much flatter rotational evolution on the early MS compared to fast rotators, because angular momentum hidden in the core is slowly transferred back to the convective envelope. Hence, the different {\it shape} of the gyrotracks computed above for slow/median rotators on the one hand and fast rotators on the other mainly arises from the differing timescale for angular momentum redistribution in the stellar interior. Indeed, given the adopted braking law, we did not find any other combination of model parameters (disk lifetime, core-envelope coupling timescale, wind braking scaling) that would reproduce the observations.

Another, yet more marginal difference between slow/median and fast rotator models is the scaling coefficient of the wind braking law ($K_1$ in Table~\ref{param}). Because we demand all models to fit the solar surface velocity at the Sun's age, this results in a scaling constant that is slightly larger for slow/median rotators than for fast ones, with $K_1$=1.8 and 1.7, respectively.  We speculate that the marginally higher braking efficiency for slow/median rotators compared to fast ones may be related to the changing topology of the surface magnetic field of solar-type stars as a function of rotation rate. Solar-type magnetospheres are known to be more organized on the large-scale in slowly rotating solar-type stars than in fast rotating ones \citep{Petit08} and hence more efficient for wind braking. However, the difference in the scaling constant of the braking law between the slow and fast rotator models amounts to a mere 5\% and is entirely driven by the condition that all the models precisely fit the surface velocity of the Sun. Since mature solar-type stars appear to exhibit some spread in their rotation rate \citep{Basri2011,Affer2012,Harrison2012}, a relaxed boundary condition at the age of the Sun would possibly erase this subtle difference in the scaling of the braking law between models.

\subsection{Wind braking and lithium depletion} 
\label{lithium}

\begin{figure}  
\centering 
\includegraphics[angle=-90,width=9cm]{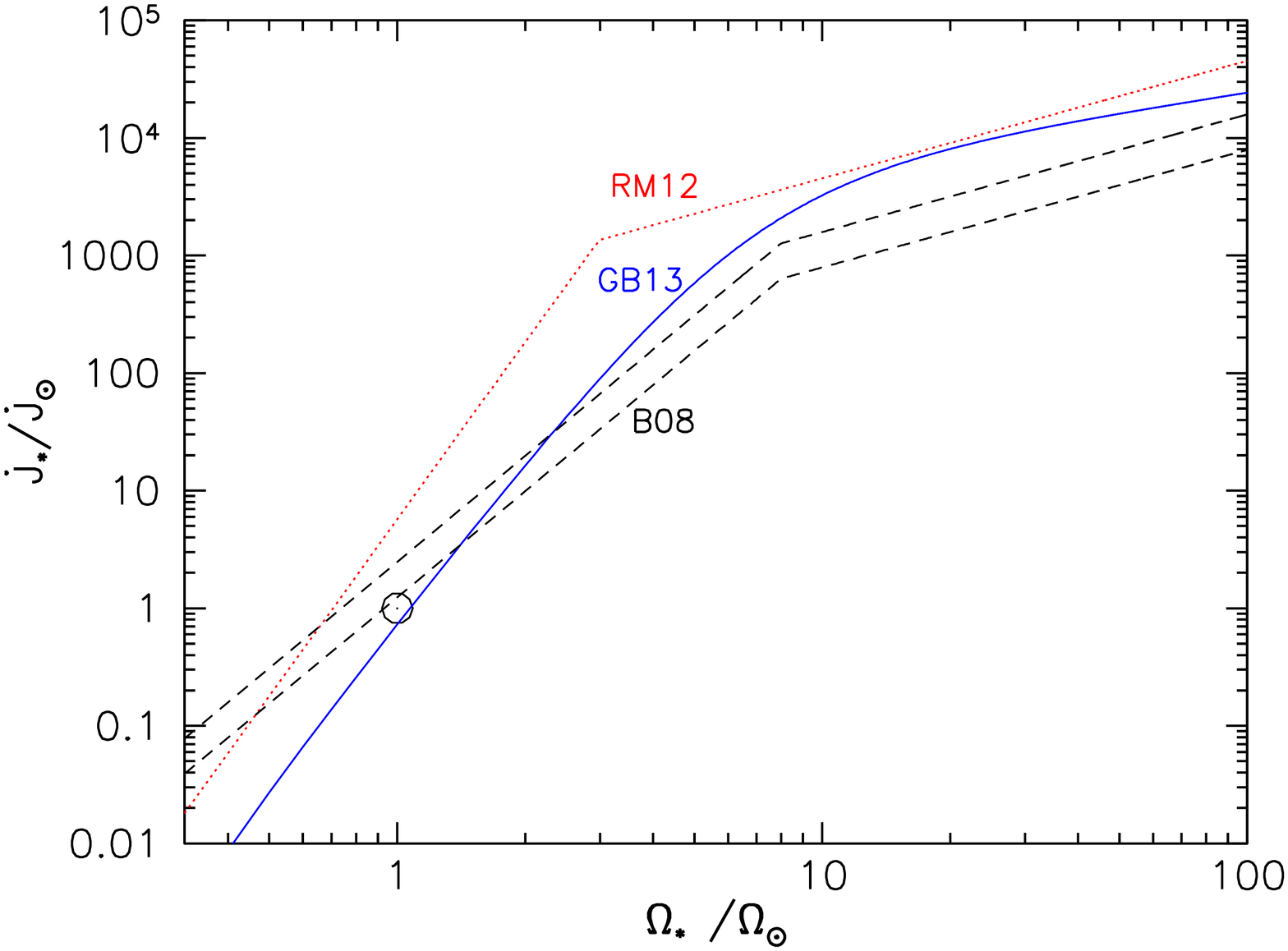}
\caption{Comparison of the angular momentum loss rate predicted by the wind braking prescriptions used in this study (solid line), in \citet{Reiners2012} (dotted line), and in \citet{Bouvier08} (dashed lines). The angular momentum loss rate is scaled to the angular momentum loss rate of the present Sun, taken to be $\dot{J}_{\odot} = 7.169 \times 10^{30} ~ \mathrm{g.cm^2.s^{-2}}$. The scaling of the braking law is  $K_1 = 1.8$ for the prescription used in this study. The Kawaler's scaling constant was $K_w = 7.5\times 10^{47}$ for slow rotators and $K_w = 3.75\times 10^{47}$ for fast rotators. Other parameters of the braking laws are summarized in Table~\ref{compbrake}.}
\label{braking}
\end{figure}

Since the rotational evolution of solar-type stars on the main sequence is primarily driven by wind braking at the stellar surface, the adopted braking law is a critical parameter of angular momentum evolution models. The models presented here implement the latest results regarding the expected properties of solar-type winds as derived from numerical simulations by \citet{Matt12} and \citet{Cranmer11}. The resulting braking law differs from the \citet{Kawaler88} prescription used in most recent modeling efforts \citep[e.g.,][]{Bouvier08,Irwin09a,Den10,Spada11} as well as from the modified Kawaler prescription proposed by \citet{Reiners2012}. We therefore proceed to discuss the comparison of our new models with those previous attempts to highlight their similarities and differences. We illustrate the different braking laws in Fig.~\ref{braking} where the angular momentum loss rate is plotted as a function of surface rotation rate\footnote{To allow for a meaningful comparison, the stellar parameters are kept constant, i.e., $R_* = 1~R_{\odot}$, $M_* = 1~M_{\odot}$ , $L_* = 1~L_{\odot}$ in the three braking laws shown in Fig.~\ref{braking} (unlike in the lower panel of the Fig.~\ref{mdotbstarjdot}). { Hence, the comparison is strictly valid only on the main sequence.}}. The main parameters and assumptions of these braking laws are summarized in Table~\ref{compbrake}.

For rotation rates between that of the Sun and a hundred times the solar value, it is seen that the prescription we use here is intermediate between those adopted by \citet{Kawaler88} and \citet{Reiners2012}, respectively. While the latter study stresses the dependency of the braking law on stellar parameters such as mass and radius, this does not come into play here, at least on the main sequence,  as we are dealing only with solar-mass stars. During the PMS, the steeper dependency of the \citet{Reiners2012} wind loss law on stellar radius will yield even stronger braking than illustrated in Fig.~\ref{braking} where the braking rate in our models is shown to be weaker than that assumed in the \citet{Reiners2012} models, especially at low velocities where the angular momentum loss rate of today's Sun appears to be overestimated by a factor of about 6. The larger angular momentum loss rate at slow rotation required by the \citet{Reiners2012} models compared to ours most likely stems from the fact that they do not allow for core-envelope decoupling but only consider solid-body rotation. For fast rotators in the saturated regime ($\Omega/\Omega_\odot\geq 10$), the two braking laws predict relatively similar angular momentum loss rates within a factor of $\sim$2.

The scaling of Kawaler's prescription in the \citet{Bouvier08} models predicts an angular momentum loss rate for the Sun ($\dot{J}(\Omega_\odot)/\dot{J_\odot}\simeq 1.2$) that is about twice as large as the solar angular momentum loss rate predicted by the models presented here ($\dot{J}(\Omega_\odot)/\dot{J_\odot}\simeq 0.7$). However, the braking rate in the present models increases more steeply than Kawaler's in the unsaturated regime (cf. Table~\ref{compbrake}) and the braking efficiency thus becomes larger than Kawaler's as soon as the angular velocity exceeds the solar value. As a result, shorter disk lifetimes are required for fast rotator models (2.5~Myr here compared to 5~Myr in the \citet{Bouvier08} models), in order to account for large velocities at the ZAMS. The shorter duration of the star-disk interaction in fast rotators is additionally supported by the newly available h~Per dataset at 13~Myr  \citep[][see Fig. \ref{model} in this paper]{Moraux}. 

In addition, as the current models and those presented in \citet{Bouvier08} have similar core-envelope coupling timescales for fast rotators (10 and 12~Myr, respectively), the more efficient braking of the outer envelope also results in enhanced differential rotation at the core-envelope boundary in the present models. While the \citet{Bouvier08} models predicted a significantly larger amount of differential rotation in slow rotators than in fast ones, the new models presented here suggest that the magnitude of core-envelope decoupling is similar in both slow and fast rotators. However, as shown in Fig. \ref{sddww}, differential rotation culminates at the ZAMS ($\simeq$ 40~Myr) for slow rotators while strong core-envelope decoupling occurs much later, at about 200~Myr, for fast rotators. Hence, even though all the models presented here do exhibit a similar level of differential rotation at some point in their evolution, their detailed rotational history may still hace an impact on lithium depletion in the long term, as discussed in \citet{Bouvier08}. It is important to emphasize, as shown by the comparison of the models presented here with previous studies, that the amount of internal differential rotation predicted by these models is quite sensitive to the adopted braking law. Robust inferences regarding the history of lithium depletion in solar-type stars would benefit from a more physical modeling of the processes involved \citep[e.g.,][]{Charbonnel05,Baraffe2010,Nascimento2010,Eggenberger10,Eggenberger12}.

\subsection{Disk lifetimes and the evolution of rotational distributions}
\label{proto}

The models presented here do not attempt to reproduce the evolution of the overall rotational distributions \citep[cf.][]{Spada11} but merely illustrate gyrotracks for slow, median, and fast rotators. Since rotation at ZAMS is determined by a set of three {\it a priori} independent parameters, the initial period $P_{init}$, the disk lifetime $\tau_{disk}$, and to a lesser extent the coupling time-scale $\tau_{c-e}$ (cf. Figs.~\ref{model} and \ref{taudec}), some degeneracy may occur between the gyrotracks. For instance, the same rotation rate can be achieved at the ZAMS by a model assuming a long initial period and a short disk lifetime and by a different model starting from a shorter initial period but assuming a longer disk lifetime. Thus, to some extent, fast rotation at ZAMS can either be reached by an initially fast rotating protostar that interact with its disk for a few Myr (the fast gyrotrack above) or by an initially slowly rotating protostar that promptly decouples from its disk. To assess whether the three gyrotracks computed above do reflect the evolution of the whole period distributions, we have to call for additional constraints. One of these is the distribution of disk lifetimes for young solar-type stars. Infrared excess and disk accretion measurements indicate that nearly all stars are born with a disk, that the disk fraction decreases to about 50\% by an age of 3~Myr, and only a small proportion of stars are still surrounded by a disk at an age of 10~Myr \citep[e.g.,][]{Hernandez08,Wyatt08,Williams11}. 

\begin{figure}  
\centering 
\includegraphics[angle=0,width=9cm]{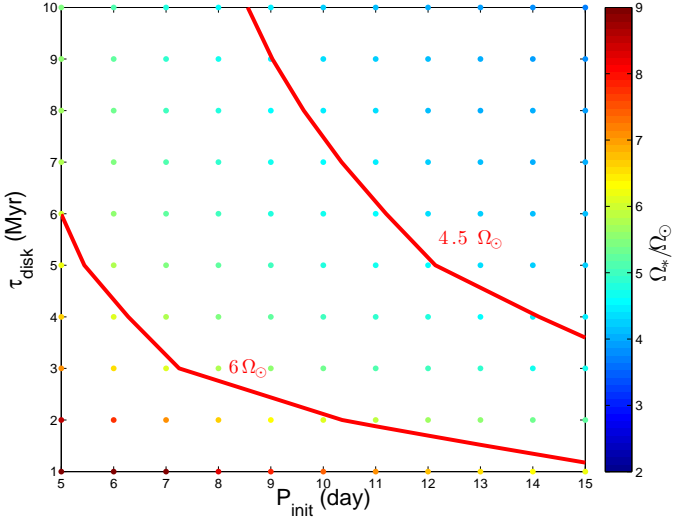}
\caption{Angular velocity at the ZAMS ($\sim$ 40 Myr) for a solar-mass star as a function of the initial period and disk lifetime in the case of the slow rotator model (cf. Table~\ref{param}). The angular velocity is scaled to the angular velocity of the present Sun, the initial period is expressed in days, and the disk lifetime in Myr.}
\label{slowzams}
\end{figure}

Combining the distribution of disk lifetimes with the distribution of initial rotational periods, we can therefore attempt to identify the PMS progenitors of slow, median, and fast rotators on the ZAMS in the framework of our models. Figure~\ref{slowzams} shows the velocity a solar-mass star will have on the ZAMS as a function of its initial period and disk lifetime, assuming the model parameters for slow rotators (cf. Table~\ref{param}). The median velocity of solar-type stars in ZAMS clusters is 6 $\Omega_\odot$ and that of slow rotators is 4.5 $\Omega_\odot$ (cf. Table~\ref{opencluster}). These values of ZAMS velocities are illustrated in Fig. ~\ref{slowzams} for initial periods ranging from 5 to 15 days and disk lifetimes between 1 and 10 Myr. The earliest stellar clusters (e.g., ONC) indicate an initial median period of 7~days, i.e., half of the stars at $\sim$1~Myr have periods of 7 days or longer. It is seen from Fig.~\ref{slowzams} that these stars will reach the median velocity of 6 $\Omega_\odot$ at ZAMS for a disk lifetime of $\sim$3~Myr or less, which is indeed the median of the disk lifetime distribution. Hence, it is fully consistent to assume that the vast majority of the 50\% of the stars that rotate most slowly at ZAMS are the evolutionary offspring of the protostars that rotate most slowly and which dissipated their disk in $\leq$3~Myr. Similarly, the slowest 25\% of the stars in PMS clusters have a period of 10~days or more. It is seen from Fig.~\ref{slowzams} that these stars have to retain their disk for $\leq$7~Myr in order to reproduce the 25\% of the slowest rotators at the ZAMS ($\Omega\leq 4.5~\Omega_\odot$). Starting with shorter initial periods, i.e., initially rotators that spin more quickly, would require significantly longer disk lifetimes to reach the same ZAMS velocities, which would then conflict with current statistical estimates of disk lifetimes. Hence,  even though there may be some degeneracy between initial periods and disk lifetimes in the modeling of the evolution of rotational distributions, our analysis suggests that most stars do follow gyrotracks qualitatively similar to those described by the slow, median, and fast rotator models presented here. 

Furthermore, as initially fast rotators have to dissipate their disks earlier in order to reach high ZAMS values (cf. Fig~\ref{model}), a trend seems to emerge for a correlation between the initial stellar velocity and PMS disk lifetime: in order to project the PMS rotational distributions onto the ZAMS in qualitative agreement with the observations, one has to assume that initially slow rotators have statistically longer-lasting disks than fast ones. A possible interpretation of this relationship is that slower rotators have more massive protostellar disks which thus dissipate on longer timescales during the PMS. In turn, this opens the intriguing possibility that the initial rotation rates result at least in part from star-disk interaction during the protostellar stage, with more massive disks being more efficient at spinning down protostars in the embedded phase \citep[e.g.,][]{FPA00}. The observed initial dispersion of rotation rate at $<$~1~Myr would thus reflect the mass distribution of protostellar disks. Additional measurements of rotational periods and disk masses for low-mass embedded protostars would be needed to test this conjecture.

\section{Conclusions}
\label{conc}

The rotational evolution of solar-type stars from birth to the age of the Sun can be reasonably accounted for by the class of semi-empirical models presented here. In these models, the physical processes at play are addressed using simplified assumptions that either rely on observational evidence (e.g., rotational regulation during PMS star-disk interaction) or are based on recent numerical simulations (e.g., wind braking). Thus, fundamental processes such as the generation of surface magnetic fields, stellar mass loss, and angular momentum redistribution, can all be scaled back to the surface angular velocity, which allows us to compute rotational evolution tracks with a minimum number of free parameters (the disk lifetime, the core-envelope coupling timescale, the scaling of the braking law). Pending more physical models still to be developed, these simplified models appear to grasp the main trends of the rotational behavior of solar-type stars between 1~Myr and 4.5~Gyr, including PMS to ZAMS spin up, prompt ZAMS spin down, and the mid-MS convergence of surface rotation rates. The models additionally predict the amount of differential rotation to be expected in stellar interiors. We caution that these predictions are mostly qualitative, as the two-zone model employed here is a crude approximation of actual internal rotational profiles. Also, we show that the evolution of internal rotation as the star ages is quite sensitive to the adopted braking law. In spite of these limitations, one of the major implications of these models is the need to store angular momentum in the stellar core for up to an age of about 1~Gyr. The build-up of a wide dispersion of rotational velocities at ZAMS and its subsequent evolution on the early MS partly reflect this process. Finally, while we have focussed here on the modeling of solar-mass stars, we will show in a forthcoming paper that similar models apply to the rotational evolution of lower mass stars.

\begin{acknowledgements}
This study was supported by the grant ANR 2011 Blanc SIMI5-6 020 01 ``Toupies: Towards understanding the spin evolution of stars" (\url{http://ipag.osug.fr/Anr_Toupies/}). We thank our partners in the ANR project, especially S. Matt for enlightening discussions on stellar wind models, and A. Palacios, C. Charbonnel, S. Brun, P. Petit, C. Zanni, and J. Ferreira for numerous discussions and comments on an early version of this work. We thank the anonymous referee for helpful comment. We acknowledge financial support from CNRS-INSU's Programme National de Physique Stellaire.   
\end{acknowledgements}

%
%
%


\def\jnl@style{\it}
\def\aaref@jnl#1{{\jnl@style#1}}

\def\aaref@jnl#1{{\jnl@style#1}}

\def\aj{\aaref@jnl{AJ}}                   
\def\araa{\aaref@jnl{ARA\&A}}             
\def\apj{\aaref@jnl{ApJ}}                 
\def\apjl{\aaref@jnl{ApJ}}                
\def\apjs{\aaref@jnl{ApJS}}               
\def\ao{\aaref@jnl{Appl.~Opt.}}           
\def\apss{\aaref@jnl{Ap\&SS}}             
\def\aap{\aaref@jnl{A\&A}}                
\def\aapr{\aaref@jnl{A\&A~Rev.}}          
\def\aaps{\aaref@jnl{A\&AS}}              
\def\azh{\aaref@jnl{AZh}}                 
\def\baas{\aaref@jnl{BAAS}}               
\def\jrasc{\aaref@jnl{JRASC}}             
\def\memras{\aaref@jnl{MmRAS}}            
\def\mnras{\aaref@jnl{MNRAS}}             
\def\pra{\aaref@jnl{Phys.~Rev.~A}}        
\def\prb{\aaref@jnl{Phys.~Rev.~B}}        
\def\prc{\aaref@jnl{Phys.~Rev.~C}}        
\def\prd{\aaref@jnl{Phys.~Rev.~D}}        
\def\pre{\aaref@jnl{Phys.~Rev.~E}}        
\def\prl{\aaref@jnl{Phys.~Rev.~Lett.}}    
\def\pasp{\aaref@jnl{PASP}}               
\def\pasj{\aaref@jnl{PASJ}}               
\def\qjras{\aaref@jnl{QJRAS}}             
\def\skytel{\aaref@jnl{S\&T}}             
\def\solphys{\aaref@jnl{Sol.~Phys.}}      
\def\sovast{\aaref@jnl{Soviet~Ast.}}      
\def\ssr{\aaref@jnl{Space~Sci.~Rev.}}     
\def\zap{\aaref@jnl{ZAp}}                 
\def\nat{\aaref@jnl{Nature}}              
\def\iaucirc{\aaref@jnl{IAU~Circ.}}       
\def\aplett{\aaref@jnl{Astrophys.~Lett.}} 
\def\apspr{\aaref@jnl{Astrophys.~Space~Phys.~Res.}}
\def\bain{\aaref@jnl{Bull.~Astron.~Inst.~Netherlands}} 
\def\fcp{\aaref@jnl{Fund.~Cosmic~Phys.}}  
\def\gca{\aaref@jnl{Geochim.~Cosmochim.~Acta}}   
\def\grl{\aaref@jnl{Geophys.~Res.~Lett.}} 
\def\jcp{\aaref@jnl{J.~Chem.~Phys.}}      
\def\jgr{\aaref@jnl{J.~Geophys.~Res.}}    
\def\jqsrt{\aaref@jnl{J.~Quant.~Spec.~Radiat.~Transf.}}
\def\memsai{\aaref@jnl{Mem.~Soc.~Astron.~Italiana}}
\def\nphysa{\aaref@jnl{Nucl.~Phys.~A}}   
\def\physrep{\aaref@jnl{Phys.~Rep.}}   
\def\physscr{\aaref@jnl{Phys.~Scr}}   
\def\planss{\aaref@jnl{Planet.~Space~Sci.}}   
\def\procspie{\aaref@jnl{Proc.~SPIE}}   

\let\astap=\aap
\let\apjlett=\apjl
\let\apjsupp=\apjs
\let\applopt=\ao

\bibliographystyle{aa}
\bibliography{Bib}

\appendix 

\section{Cluster parameters}
\label{clustparam}

In this section we detail the parameters of the open clusters and star forming regions whose rotational distributions we used to constrain our model simulations. Table~\ref{openclusterparam} summarizes their properties. 

\begin{table*}[ht!]
\caption{Cluster parameters.}             
\label{openclusterparam}      
\centering                          
\begin{tabular}{c c c c c c c}       
\hline\hline                
Cluster & Age & Ref. & Metallicity & Ref. & Distance & Ref. \\  
 & (Myr)& & ([Fe/H]) &  & (pc) &  \\
\hline    
ONC & 0.8, 2 & 1,2 & -0.01$\pm$0.04 & 3 & 450, 470$\pm$70 & 1,2  \\
NGC 6530 & 1-2.3 & 4,5,6,7 & $[-0.3,0.3]$ & 5,7 & 1250 & 5,7 \\
NGC 2264 & 2-3 & 8,9,10 & -0.09$\pm$0.3, -0.16 & 11 & 750-950 & 12,13 \\
NGC 2362 & 3, $5^{+1}_{-2}$ & 14,15 & (solar?) & 16 & 1500 & 15,16 \\
h PER & $14 \pm 1$ & 17 &  (Z = 0.019) & 17 & $2079-2290^{+87}_{-82}$  & 17,18 \\
NGC 2547 &  $38.5^{+3.5}_{-6.5}$ & 19,20 & $ [-0.21,-0.12]$ & 21 & 361-457 & 18  \\
Pleiades & 120-125 & 22 & $0.03 \pm 0.05$ & 23 & 133 & 24 \\
M 50 & 130 & 25 & (solar?) & 25,26 & $1000^{+81}_{-75}$ & 25  \\
M 35 & 150-180 & 25,27 & $-0.21\pm0.1$ & 28 & $912^{+70}_{-55}$ & 25   \\
M 37 & $550 \pm 30$ & 29 & $0.045\pm0.044$ & 29 & $1383-1490\pm120$ & 29,30  \\
Praesepe & $578\pm12$ & 31 & [$0.038\pm0.039$,$0.27\pm0.10$] & 32,33 & $182\pm6-187$ & 34,35 \\
Hyades & $628\pm14$ & 31 & $0.14\pm0.05$ & 36 & $46.45\pm0.5$ & 35  \\
NGC 6811 & $1000\pm170$ & 37 & -0.19 & 37 & $1106^{+95}_{-88}-1240$ & 34,37  \\
\hline   
\end{tabular}
\tablebib{(1)~\citet{Herbst02}; (2) \citet{Hillenbrand97}; 
(3) \citet{O'Dell00}; (4) \citet{Henderson11}; (5) \citet{Prisinzano05}; (6) \citet{Prisinzano07}; (7) \citet{Prisinzano12}; (8) \citep{Sung09}; (9) \citet{Teixeira12}; (10) \citet{Affer2013}; (11) \citet{Tadross03}; (12) \citet{Flaccomio99}; (13) \citet{Mayne08}; (14) \citet{Mayne07}; (15) \citet{Irwin08a}; (16) \citet{Dahm07}; (17) \citet{Currie10}; (18) \citet{Kharchenko05}; (19) \citet{Irwin08b} ; (20) \citet{Naylor06}; (21) \citet{Paunzen10} ; (22) \citet{Stauffer98} ; (23) \citet{Soderblom98}; (24) \citet{Soderblom05}; (25) \citet{Kalirai03}; (26) \citet{Irwin09b}; (27) \citet{vonHippel02}; (28) \citet{Barrado01} ; (29) \citet{Hartman08a}     ; (30) \citet{Wu09}; (31) \citet{Delorme11}; (32) \citet{Friel92}; (33) \citet{Pace08}; (34) \citet{Kharchenko05}; (35) \citet{vLeeuwen09} ; (36) \citet{Perryman98} ; (37) \citet{Janes13}.}
\end{table*}

\subsection{ONC}

The Orion Nebula Cluster is a very young cluster, with an age of 0.8-2 Myr \citep{Herbst02,Hillenbrand97} and located at a distance of about 450 pc \citep{Herbst02,Hillenbrand97}. The rotational data used in this study come from \citet{Herbst02} and the mass estimates are from \citet{Hillenbrand97} who derived them using the \citet{D'Antona94} isochrone models. The metallicity of the ONC is $[Fe/H] = -0.01 \pm 0.04$ \citep{O'Dell00}. 


\subsection{NGC 6530}

The age of NGC 6530 lies between 1 and 2.3 Myr \citep{Prisinzano05,Mayne07,Henderson11} and its distance is about 1250 pc \citep{Prisinzano05,Prisinzano12}. The rotational data used here come from \citet{Henderson11}. Stellar masses were estimated by \citet{Prisinzano05,Prisinzano07,Prisinzano12} by interpolating the theoretical tracks and isochrones of \citet{Siess00} to the stars location in the V vs. V-I color-magnitude diagram. \citet{Prisinzano05} assumed a solar metallicity and used the \citet{Siess00} models with Z = 0.02, Y = 0.277, X = 0.703. In \citet{Prisinzano12} a metallicity range of $-0.3 < [Fe/H] < 0.3$ is considered.

\subsection{NGC 2264}

The NGC 2264 cluster is 2-3 Myr old \citep{Sung09,Teixeira12,Affer2013} located at a distance between 750 and 950 pc \citep{Flaccomio99,Mayne08,Baxter09,Cauley12,Affer2013}. The rotational data used here, as well as mass estimates, come from \citet{Affer2013} who used the V vs. V-I CMD together with the \citet{Siess00} isochrones to derive stellar masses.  NGC2264 has a metallicity estimated to range from solar to slightly metal-poor \citep{Tadross03,Cauley12}.

\subsection{NGC 2362}

The age of NGC 2362 is about $3-5^{+1}_{-2}$ Myr \citep{Moitinho01,Mayne07,Irwin08a} and the cluster is located at a distance of about 1500 pc \citep{Moitinho01,Dahm07,Irwin08a}. The rotational data used here come from \citet{Irwin08a} as well as the masses estimates. They used the I magnitude together with the \citet{Baraffe98} 5 Myr isochrones to derive stellar masses. 
\citet{Dahm07} assumed a solar-metallicity for this cluster.

\subsection{h PER}

The h PER (NGC869) clsuter is $14\pm1$ Myr old \citep{Currie10} located at a distance of about 2.1~kpc \citep{Kharchenko05,Currie10}. The rotational data and mass estimates used in this study come from \citet{Moraux}. They used the I magnitude together with the \citet{Siess00} 13.8 Myr isochrone with an extinction $A_I = 1$~mag \citep{Currie10} to derive stellar masses. \citet{Currie10} reported a metallicity $Z = 0.019$.

\subsection{NGC 2547}

The age of NGC 2547 is about $38.5^{+3.5}_{-6.5}$ Myr \citep{Naylor06,Irwin08b} and it lies at a distance of $361^{+19}_{-8}-457$ pc \citep{Kharchenko05,Naylor06}. The rotational periods and stellar masses used here come from \citet{Irwin08b}, who used the I magnitude together with the \citet{Baraffe98} 40 Myr isochrones to determine the masses. The reddening corresponds to $A_V = 0.186$ \citep{Naylor06}. \citet{Paunzen10} report sub-solar metallicity $-0.21 < [Fe/H] < -0.12$.

\subsection{Pleiades}

The Pleiades is a 120-125 Myr old cluster \citep{Stauffer98} situated at 133 pc \citep{Soderblom05}. The rotational data used in this study come from \citet{Hartman10} as do the mass estimates. They used the $M_K$ magnitude together with the Yonsei-Yale (Y2) isochrones \citep{Yi01} with an extinction $A_K = 0.01~mag$ \citep{Stauffer07} to determine the masses. By using the \citet{Siess00} 125 Myr isochrones models we recalculated the mass of the stars and we found that the ones derived with the \citet{Siess00} models are higher by $5\%$ at most. \citet{Soderblom98} report a metallicity of $[Fe/H] = 0.03 \pm 0.05$.

\subsection{M 50}

The age of M50 (NGC 2323) is about $130$ Myr \citep{Kalirai03,Irwin09b} and its distance $1000^{+81}_{-75}$ pc \citep{Kalirai03}. The rotational  data and mass estimates used here come from \citet{Irwin09b} who used the I magnitude together with the \citet{Baraffe98} 130 Myr isochrones to determine the masses. The reddening of the cluster is $E(B-V)=0.22$ mag corresponding to $A_V = 0.68$ \citep{Kalirai03}. The metallicity of M50 is believed to be solar \citep{Kalirai03,Irwin09b}.

\subsection{M 35}

The age estimate for M35 (NGC 2168) ranges from $150$ Myr \citep{vonHippel02,Meibom09} to 180 Myr \citep{Kalirai03} and its distance is $912^{+70}_{-65}$ pc \citep{Kalirai03}. The rotational data used here come from \citet{Meibom09}. We used the \citet{Siess00} 130 Myr isochrones together with the $(B-V)_0$ measurements from \citet{Meibom09} to estimate the stellar masses. The reddening of the cluster is $E(B-V)=0.20$ mag, corresponding to $A_V = 0.62$ \citep{Kalirai03}. The metallicity of M35 is $[Fe/H] = -0.21\pm0.10$ ($Z = 0.012$) \citep{Barrado01,Kalirai03}.

\subsection{M 37}

The age of M37 (NGC 2099) is about $550\pm30$ Myr \citep{Hartman08a} and its distance $1383-1490\pm120 $ pc \citep{Hartman08a,Wu09}. The rotational data used here come from \citet{Hartman09}. We used the \citet{Siess00} 550 Myr isochrones together with the I magnitude measurements from \citet{Hartman09} to estimate the mass of the stars. The reddening of the cluster is $E(B-V)=0.227 \pm 0.038$ mag, corresponding to $A_V = 0.70$ \citep[assuming $R_V = 3.1$,][]{Hartman08a} and $A_I = 0.852$. The metallicity of M37 has been estimated to be $[Fe/H] = 0.045\pm0.044$ and $0.09$ \citep{Hartman08a,Wu09} and $Z = 0.011 \pm 0.001$ - $0.019$ \citep{Kalirai05,Kang07}.

\subsection{Praesepe}

Praesepe (M44, NGC2632) is a $578\pm12$ Myr old cluster \citep{Delorme11} located at a distance of $182\pm 6-187$ pc \citep{Kharchenko05,vLeeuwen09}. The rotational data used here come from \citet{Delorme11}. We used the \citet{Siess00} 578 Myr isochrones together with the $(J-K)$ measurements from \citet{Delorme11} to estimate the mass of the stars. The reddening of the cluster is $E(B-V)=0.027 \pm 0.004$ mag, corresponding to $A_V = 0.083$ \citep{Taylor06} and $E(J-K)=0.012$. Metallicity estimates range from $[Fe/H] = 0.038\pm0.039$ to $0.27\pm0.10$ \citep{Friel92,Pace08}.

\subsection{Hyades}

The Hyades (Melotte 25) is a $628\pm14$ Myr old cluster \citep{Delorme11} located at a distance of $46.45\pm0.5$ pc \citep{vLeeuwen09}. The rotational data used here come from \citet{Delorme11}. We used the \citet{Siess00} 625 Myr isochrones together with the $(J-K)$ measurements from \citet{Delorme11} to estimate the mass of the stars. The reddening of the cluster is negligible, $E(B-V) \leq 0.0014$ mag \citep{Taylor06}. The metallicity of the Hyades is $[Fe/H] = 0.14\pm0.05$ \citep{Perryman98}.

\subsection{NGC 6811}

The cluster NGC6811 is $1000\pm170$ Myr old \citep{Kharchenko05,Meibom2011,Janes13} located  at a distance of $1106^{+95}_{-88}-1240$ pc \citep{Kharchenko05,Janes13}. The rotational data used here come from \citet{Meibom2011}. We used the \citet{Siess00} 1 Gyr isochrones together with the 2MASS J, H, and K measurements from the Kepler archives (\url{http://archive.stsci.edu/kepler/kepler_fov/search.php}) to estimate the mean mass of the stars adopting a distance of 1240 pc. The reddening of the cluster is $E(B-V)=0.074 \pm 0.024$ mag, corresponding to $A_V = 0.23$ \citep{Janes13}. The metallicity of NGC6811 is $[Fe/H] = -0.19$ and $Z = 0.012\pm0.004$ \citep{Janes13}.

\end{document}